\documentclass[13 pt]{article}
\usepackage{amssymb}

\def\acal{{\Large{\textit{a}}}}
\def\bcal{{\Large{\textit{b}}}}
\def\ccal{{\Large{\textit{c}}}}
\def\acalt{\tilde{\Large{\textit{a}}}}
\def\bcalt{\tilde{\Large{\textit{b}}}}
\def\ccalt{\tilde{\Large{{\textit{c}}}}}

\newcommand{\beq}{\begin{equation}}
\newcommand{\eeq}{\end{equation}}
\newcommand{\bea}{\begin{eqnarray*}}
\newcommand{\eea}{\end{eqnarray*}}
\newcommand{\beqa}{\begin{eqnarray}}
\newcommand{\eeqa}{\end{eqnarray}}

\begin{document}

\newfont{\elevenmib}{cmmib10 scaled\magstep1}%

\newcommand{\Title}[1]{{\baselineskip=26pt \begin{center}
            \Large   \bf #1 \\ \ \\ \end{center}}}
\hspace*{2.13cm}%
\hspace*{1cm}%
\newcommand{\Author}{\begin{center}\large
           Pascal Baseilhac\footnote{
baseilha@phys.univ-tours.fr} 
\end{center}}
\newcommand{\Address}{{\baselineskip=18pt \begin{center}
           \it Laboratoire de Math\'ematiques et Physique Th\'eorique CNRS/UMR 6083,\\
           F\'ed\'eration Denis Poisson,\\
Universit\'e de Tours, Parc de Grandmont, 37200 Tours, France
      \end{center}}}
\baselineskip=13pt

\bigskip
\vspace{-1cm}

\Title{The $q-$deformed analogue of the Onsager algebra:\\ Beyond the Bethe ansatz approach}\Author

\vspace{- 3mm}
 \Address

\vskip 0.3cm

\centerline{\bf Abstract}\vspace{0.3mm}  \vspace{1mm}
The spectral properties of operators formed from generators of the $q-$Onsager non-Abelian infinite dimensional algebra are investigated. 
Using a suitable functional representation, all eigenfunctions are shown to obey a second-order $q-$difference equation (or its degenerate discrete version).
In the algebraic sector associated with polynomial eigenfunctions (or their discrete analogues), Bethe equations naturally appear. Beyond this sector, where the Bethe ansatz approach is not applicable in related massive quantum integrable models, the eigenfunctions are also described.  
The spin-half XXZ open spin chain with general integrable boundary conditions is reconsidered in light of this approach: all the eigenstates are constructed. In the algebraic sector which corresponds to special relations among the parameters, known results are recovered.\vspace{2mm}



\section{Introduction}
In the context of quantum integrable systems  (continuum or lattice), finding the non-Abelian infinite dimensional algebra responsible of the existence of an (in)finite number of mutually commuting conserved quantities is a challenge. Indeed, the success of conformal field theory essentially relies on the Virasoro algebra and its representation theory which allows to derive exact results such as the energy spectrum and exact correlation functions \cite{BPZ84}.
For lattice models such as the Ising \cite{Ons}, XY \cite{DG}, superintegrable ${\mathbb Z}_n-$chiral Potts model \cite{Potts} and some generalizations \cite{Ahn}, it is the Onsager non-Abelian Lie algebra \cite{Ons} which plays an analogous role and allows to derive the exact spectrum for any conserved quantity of the model using solely its representation theory \cite{Davies}. In these latter models, the transfer matrix can be decomposed on a basis of mutually commuting conserved quantities $I_{2k+1}$, $k=0,1,...$ which form an Abelian subalgebra of the Onsager algebra. In particular, each quantity $I_{2k+1}$ is expressed in terms of certain nonlinear combinations of the two fundamental operators $A_0,A_1$ which satisfy the Dolan-Grady relations \cite{DG} 
\beqa
[A_0,[A_0,[A_0,A_1]]]=\rho_0[A_0,A_1]\qquad \mbox{and}\qquad [A_1,[A_1,[A_1,A_0]]]=\rho_0[A_1,A_0]\ ,\label{relDG}
\eeqa
and generate the Onsager algebra with generators $A_k,G_k$, $k$ integer \cite{Perk,Davies}. Note that the  parameter $\rho_0$ as well as the realizations of $A_0,A_1$ are model-dependent: for the Ising model $\rho_0=16$ and for the superintegrable ${\mathbb Z}_n-$chiral Potts model $\rho_0=n^2$. In both models, the generators are realized in terms of the $sl_2$ loop algebra and the solution of the spectral problem for the transfer matrix or any conserved quantity has been solved using solely the representation theory of (\ref{relDG}) \cite{Davies}.\vspace{1mm}
  
In the last few years, there has been some renewed interest in the Onsager algebra, Dolan-Grady integrable structure \cite{Date00,Gehl1,Plyu02,Gehl,Ro05,Fair} and its deformation \cite{qDG,TriDiag,qOns} in the context of solvable lattice models  as well as in the representation theory of orthogonal polynomials \cite{Gehl1,Ter01,Ter03,Cheb,TDpair}. Related with the subject of this paper, a $q-$deformed analogue of the Onsager algebra has been exhibited in various quantum integrable systems (Azbel-Hofstadter model, XXZ open spin chain with the most general boundary conditions \cite{qOns},...) which are usually studied using the quantum inverse scatterring method. 
As shown in \cite{qDG,TriDiag} and similarly to the undeformed case \cite{Ons,DG}, in these models the integrability property is actually related with the existence of two operators ${\textsf W_0},{\textsf W_1}$ which satisfy the {\it tridiagonal relations} (also called the $q-$deformed Dolan-Grady relations) with scalars $\rho,\rho^*$
\beqa
[\textsf{W}_0,[\textsf{W}_0,[\textsf{W}_0,\textsf{W}_1]_q]_{q^{-1}}]=\rho[\textsf{W}_0,\textsf{W}_1]\
,\qquad
[\textsf{W}_1,[\textsf{W}_1,[\textsf{W}_1,\textsf{W}_0]_q]_{q^{-1}}]=\rho^*[\textsf{W}_1,\textsf{W}_0]\ .
\label{qDG}  \eeqa
More generally, they generate a $q-$deformed analogue of the Onsager algebra with generators $\{{\textsf W}_{-k},{\textsf W}_{k+1}$, ${\textsf G}_{k+1},{\tilde{\textsf G}}_{k+1}\}$ introduced and studied in details in \cite{qOns}. In addition, it has been discovered that the transfer matrix in these models - calculated in the inverse scattering framework \cite{Skly88} - can be simply written in the form
\beqa
t(u)= \sum_{k=0}^{N-1}{\cal F}_{2k+1}(u)\ {{\cal I}}_{2k+1} + {\cal F}_0(u)\ I\!\!I\ ,\label{tfin}
\eeqa
where ${\cal F}_{2k+1}(u)$ are certain rational functions of the spectral parameter $u$ and $\{{{\cal I}}_{2k+1}\}$, $k=0,1,..,N-1$, is a family of mutually commuting operators acting on the quantum space of the system. Remarkably, these operators form an ordered subset\,\footnote{In the case of finite dimensional representations, the infinite hierarchy truncates due to the existence of linear relations among the generators. For the XXZ open spin chain, see \cite{qOns}.} of the $q-$deformed analogue of the Dolan-Grady infinite hierarchy \cite{qDG,TriDiag}. Written in terms of the generators of the infinite dimensional $q-$deformed analogue of the Onsager algebra, this hierarchy takes the form: 
\beqa
{\cal I}_{2k+1}=\kappa {\textsf W}_{-k} + \kappa^*{\textsf W}_{k+1} + \frac{\kappa_+}{k_+}{\textsf G}_{k+1} 
+ \frac{\kappa_-}{k_-} {\tilde{\textsf G}}_{k+1}\ , \qquad k=0,1,...\label{IN}
\eeqa
for arbitrary parameters $\kappa,\kappa^*,\kappa_\pm,k_\pm$. Although technically rather lengthy, by analogy with the undeformed case \cite{DG} it is alternatively possible to write the operators $\{{\textsf W}_{-k},{\textsf W}_{k+1},{\textsf G}_{k+1},{\tilde{\textsf G}}_{k+1}\}$ for $k\geq 0$ in terms of nonlinear combinations of the fundamental ones $\{{\textsf W}_{0},{\textsf W}_{1}\}$. For instance, one has \cite{TriDiag} (see also \cite{qOns})
\beqa
{\cal I}_{1}=\kappa {\textsf W}_{0} + \kappa^*{\textsf W}_{1} + \frac{\kappa_+}{k_+}\big[{\textsf W}_1,{{\textsf W}_0}\big]_q 
+ \frac{\kappa_-}{k_-} \big[{\textsf W}_0,{{\textsf W}_1}\big]_q\ ,\label{I1}
\eeqa
where the $q-$commutator $[X,Y]_{q}=q^{1/2}XY-q^{-1/2}YX$ has been introduced. Note that the parameters 
$N$, $\rho,\rho^*$, $\kappa,\kappa^*$, $\kappa_\pm,k_\pm$, $q$ and the functions ${\cal F}_{2k+1}(u)$  obviously depend on the physical characteristic of the model under consideration\,\footnote{The simplest example with a transfer matrix of the form (\ref{tfin}) corresponds to $N=1$: it is the Azbel-Hofstadter model which describes the problem of Bloch electrons in a magnetic field on a two-dimensional lattice. There, the elements ${\textsf W_0},{\textsf W_1}$ are realized in terms of the Weyl algebra \cite{qDG}. One has ${\cal F}_{1}(u)=u^2$, $\kappa,\kappa^*,\kappa_\pm,k_\pm$ are related with length scales along the directions of the lattice and $q\neq 1$, related with the magnetic flux per plaquette, is a root of unity. 
An other model is the XXZ open spin chain with nondiagonal boundary conditions. In this case ${\textsf W_0},{\textsf W_1}$ are expressed in terms of $U_{q^{1/2}}(sl_2)$ operators acting on the Hilbert space of the whole spin chain with $N$ sites \cite{qDG,TriDiag}. The parameters  $\kappa,\kappa^*,\kappa_\pm,k_\pm$ are related with integrable boundary conditions (see Section 3).}.\vspace{1mm}
 
Having in mind the use of the Onsager algebra in the Ising \cite{Ons}, superintegrable ${\mathbb Z}_n-$chiral Potts model \cite{Potts} or the Virasoro algebra in conformal field theory \cite{BPZ84}, it seems to us important to analyse the models which enjoy the hidden dynamical symmetry (\ref{qDG}) from the point of view of the representation theory associated with the tridiagonal algebra (\ref{qDG}). Indeed, for several models - for instance the XXZ open spin chain with arbitrary integrable boundary conditions - the standard {\it Bethe ansatz approach} does not apply except at some special points in the boundary parameters space. As we are going to see, the analysis based on the representation theory of (\ref{qDG}) gives a clear understanding of the structure of the space of states in models with transfer matrices of the form (\ref{tfin}).\vspace{1mm}

Among the interesting problems in integrable lattice models with hidden symmetry (\ref{qDG}), one fundamental is to identify the eigenstates
of the transfer matrix, and corresponding eigenvalues. According to the decomposition (\ref{tfin}) and the fact that the spectral parameter $u$ is arbitrary, the structure of these eigenstates is encoded in the spectral problem for the $q-$deformed Dolan-Grady hierarchy $\{{\cal I}_{2k+1}\}$, $k=0,1,2,...$. By virtue of \cite{qOns}
\beqa
[{\cal I}_{2k+1},{\cal I}_{2l+1}]=0 \qquad \mbox{and}\qquad {\cal I}_{2k+1}\Psi = {\Lambda}_{2k+1}\Psi\ \label{inv}
\eeqa
for all\  $k,l=0,1,...$, it follows that the structure of the eigenstates $\Psi$ is completely determined from the spectral problem for any operator ${\cal I}_{2k+1}$, provided its eigenvalues are {\it non-degenerate}. For instance, one can focus on the simplest operator, namely ${\cal I}_1$ defined by (\ref{I1}). If its spectrum is non-degenerate (see details in further sections), finding the eigenstates of (\ref{IN}) is equivalent to solve ${\cal I}_{1}\Psi = {\Lambda}_{1}\Psi$. \vspace{1mm}
 
In this paper, we establish a direct relationship between the spectral problem for the hierarchy (\ref{IN}) and a second-order $q-$difference equation in one variable - or its discrete version - satisfied by the states $\Psi$ in a suitable functional representation.  The necessary and sufficient information to solve this spectral problem is shown to be encoded in the fundamental operators ${\textsf W_0},{\textsf W_1}$, which properties uniquely determine the eigenfunctions $\Psi(z)$. Some examples are considered in details. \vspace{1mm} 

The paper is organized as follows. In Section 2, an infinite dimensional module of the tridiagonal algebra (\ref{qDG}) is constructed. It is shown that the spectrum of the operator ${\textsf W_0}$ can be algebraized i.e. corresponds to (possibly degenerate) polynomial eigenfunctions which roots satisfy a set of Bethe equations. In this eigenbasis of ${\textsf W_0}$, the action of ${\textsf W_1}$ is described.
Then, the spectral problem for (\ref{I1}) is considered: in the algebraic sector associated with special relations among the parameters $\kappa,\kappa^*,\kappa_\pm,k_\pm\neq 0$, the operator ${\cal I}_1$ admits polynomial eigenfunctions. Beyond the algebraic sector i.e. 
for arbitrary parameters $\kappa,\kappa^*,\kappa_\pm,k_\pm\neq 0$, the (non-polynomial) eigenfunctions $\Psi(z)$ of (\ref{I1}) are expanded on the (infinite dimensional) basis of polynomials eigenfunctions of ${\textsf W_0}$. Corresponding weight functions are determined by a coupled system of recurrence relations, which coefficients follow from the fundamental operator ${\textsf W_1}$. 
In Section 3, the same approach is applied to the case of finite dimensional modules of the tridiagonal algebra. In particular, we focus on the XXZ open spin chain with integrable boundary conditions. The eigenstates $\Psi(z_s)$ of the transfer matrix, represented by functions defined on a discrete support, are shown to satisfy a set of discrete $q-$difference equations. For general values of the boundary parameters, the eigenstates are constructed. For special linear relations among the boundary parameters, a subset of eigenstates (called Bethe eigenstates) are given by polynomial eigenfunctions defined on a discrete support, which roots are solutions of Bethe equations in agreement with known results. An alternative construction of these eigenstates is also proposed in Appendix A. Comments follow in the last Section.

\section{A second-order $q-$difference equation from the infinite hierarchy}
In this Section, we construct an infinite dimensional module of the tridiagonal algebra (\ref{qDG}) on which the infinite $q-$deformed Dolan-Grady hierarchy (\ref{IN}) acts. Although the infinite dimensional modules are more complicated to study than the finite dimensional ones, in our example the main properties of the finite dimensional case (eigenvalue sequences, block diagonal structure in the dual eigenbasis and duality of the fundamental operators) hold. These properties will be used to derive a $q-$difference equation for the eigenstates of (\ref{IN}). 

\subsection{An infinite dimensional module of the tridiagonal algebra}
Let $V$ denotes an irreducible infinite dimensional module on which ${\textsf W_0},{\textsf W_1}$ are both diagonalizable. Suppose $q$ is not a root of unity.
Let $z$ denote a complex variable. Define $\eta$ to be the ``shift'' operator such that $\eta^{\pm 1}(z)=q^{\pm 1}z$.  Our aim is to endow the vector space of all functions in the variable $x\equiv z+z^{-1}$ with the module structure of a certain tridiagonal algebra (\ref{qDG}). By analogy with the finite dimensional case (see \cite{Ter01,Ter03,TDpair}), for the linear transformations ${\textsf W}_0$, ${\textsf W}_1$ it is natural to consider the following realization\,\footnote{If ${\overline \phi}(z)\neq {\phi}(z^{-1})$, we assume there exists at least one other realization of the operator ${\textsf W}_0$ which spectrum is identical.} of the fundamental operators:
\beqa
{\textsf W_0}: && \; \phi(z)\eta \;+ \;{\overline \phi}(z)\eta^{-1}\;+\;(d(z,z^{-1})-\phi(z)-{\overline \phi}(z))I\ ,
\nonumber
\\
{\textsf W_1}: && \;z+z^{-1}\ .\label{real}
\eeqa
Here, ${\phi}(z),{\overline \phi}(z),d(z,z^{-1})$ are functions of $z$ which are not necessarely unique, as the spectrum of ${\textsf W}_1$ may contain degeneracies. As we will see in Section 3, this kind of situation happens in the case of finite dimensional modules too. The existence of such degeneracies induces $dim(V^*_s)$ possible realizations of the operator ${\textsf W_0}$.\vspace{1mm}

Our purpose is now to find the necessary and sufficient conditions on ${\phi}(z),{\overline \phi}(z),d(z,z^{-1})$ such that the tridiagonal relations (\ref{qDG}) are satisfied. First, for any ${\phi}(z),{\overline \phi}(z),d(z,z^{-1})$ it is easy to check that (\ref{real}) automatically solves the second equation in (\ref{qDG}) provided one identifies
\beqa
\rho^*=-(q-q^{-1})^2\ .\label{rhostar}
\eeqa
Then, replacing (\ref{real}) in the l.h.s. of the first equation of (\ref{qDG}) a straightforward calculation gives
\beqa
[{\textsf W_0},[{\textsf W_0},[{\textsf W_0},{\textsf W_1}]_q]_{q^{-1}}]&=&\eta(\beta(\phi,{\overline \phi},d))\phi(z)\eta(\phi(z))\eta^2 - \eta^{-1}(\beta(\phi,{\overline \phi},d)){\overline \phi}(z)\eta^{-1}({\overline \phi}(z))\eta^{-2}\nonumber\\ 
&&\gamma(\phi,{\overline \phi},d)\phi(z)(\eta(x)-x)\eta + \eta^{-1}(\gamma(\phi,{\overline \phi},d)){\overline \phi}(z)(\eta^{-1}(x)-x)\eta^{-1}\ \label{lhs}
\eeqa
with
\beqa
\beta(\phi,{\overline \phi},d)&=&\big(\eta(\phi(z))+\eta({\overline \phi}(z))-\eta(d(z,z^{-1}))\big)\big((q^2-q)z+(q^{-2}-q^{-1})z^{-1}\big)\nonumber\\
&&+\big(\phi(z)+{\overline \phi}(z)-d(z,z^{-1})\big)\big((q-q^{-1})(z^{-1}-z)\big)\nonumber\\
&&+\big(\eta^{-1}(\phi(z))+\eta^{-1}({\overline \phi}(z))-\eta^{-1}(d(z,z^{-1}))\big)\big((q^{-1}-q^{-2})z+(q-q^2)z^{-1}\big)\ ,\label{beta}
\eeqa
\beqa
\gamma(\phi,{\overline \phi},d)&=& \Big(\eta(\phi(z))+\eta({\overline \phi}(z))-\eta(d(z,z^{-1}))\Big)^2 +\Big(\phi(z)+{\overline \phi}(z)-d(z,z^{-1})\Big)^2\nonumber\\
&&-(q+q^{-1})\Big(\eta(\phi(z))+\eta({\overline \phi}(z))-\eta(d(z,z^{-1}))\Big)\Big(\phi(z)+{\overline \phi}(z)-d(z,z^{-1})\Big)\nonumber\\
&& + (2+q+q^{-1})\phi(z)\eta({\overline \phi}(z)) +  {\overline \phi}(z)\eta^{-1}(\phi(z)) + \eta(\phi(z))\eta^2({\overline \phi}(z))\nonumber\\
&& + \frac{(1+q+q^{-1})}{\eta(x)-x}\Big( {\overline \phi}(z)\eta^{-1}(\phi(z)) \big(\eta^{-1}(x)-x\big)  
+ \eta(\phi(z))\eta^{2}({\overline \phi}(z)) \big(\eta(x)-\eta^2(x)\big)\Big)\ .\label{gamma}
\eeqa
On the other hand,
\beqa
[{\textsf W_0},{\textsf W_1}]=\phi(z)(\eta(x)-x)\eta + {\overline \phi}(z)(\eta^{-1}(x)-x)\eta^{-1}\ .\label{rhs}
\eeqa
Identifying (\ref{lhs}) and (\ref{rhs}), we immediately deduce that the operators ${\textsf W_0},{\textsf W_1}$ with (\ref{real}) satisfy the tridiagonal algebraic relations (\ref{qDG}) iff the functions ${\phi}(z),{\overline \phi}(z),d(z,z^{-1})$ solve the constraints:
\beqa
\beta(\phi,{\overline \phi},d)=0 \qquad \mbox{and}\qquad \gamma(\phi,{\overline \phi},d)=\rho\ .\label{cond}
\eeqa 

The equation (\ref{beta}) being a second-order $q-$difference equation with simple Laurent polynomials as coefficients, we restrict our attention to the solutions ${\phi}(z),{\overline \phi}(z)$ which are {\it rational} functions of $z$ and $d(z,z^{-1})$ is supposed to be a Laurent polynomial in $z$. In this case, define the family
\beqa
\phi(z)=\frac{1}{(1-z^2)(1-qz^2)}\frac{\prod_{k=1}^{2N+2}(1-\chi_kz)}{\prod_{k=1}^{2N-2}(1-\xi_kz)}\qquad \mbox{and}\qquad {\overline \phi}(z)=\phi(z^{-1})\ ,\label{fonction}
\eeqa
where we have introduced $4N$ scalars $\{\chi_k\}$, $\{\xi_k\}$ in an arbitrary field ${\mathbb K}$. Replacing (\ref{fonction}) in (\ref{beta}), (\ref{gamma}), eqs. (\ref{cond}) impose some constraints on the parameters:\vspace{1mm} 

For $N=1$, it is straightforward to check that the constraints (\ref{cond}) are satisfied with the identification (\ref{rhostar})
and 
\beqa 
\rho=-(q-q^{-1})^2\chi_1\chi_2\chi_3\chi_4q^{-1}\ ,\qquad d(z,z^{-1})=1+\chi_1\chi_2\chi_3\chi_4q^{-1}\ ,
\eeqa  
for any choice of the parameters $\{\chi_1,\chi_2,\chi_3,\chi_4\}$. Notice that the realization (\ref{real}) for the operator ${\textsf W}_0$ in this case coincides with the second-order Askey-Wilson $q-$difference operator \cite{AW}.\vspace{1mm}

For $N=2$, the first constraint in (\ref{cond}) is satisfied for the choice 
\beqa
d(z,z^{-1})=1+\chi_1\chi_2\chi_3\chi_4\chi_5\chi_6\xi^{-1}_1\xi_2^{-1}q^{-1}\ ,
\eeqa
whereas the second constraint yields to the following relations among the parameters:
\beqa
\sum_{k=-2}^{3} \Big(\sum_{i_1<...<i_{3+k}}(-1)^{k}\chi_{i_1}\chi_{i_2}...\chi_{i_{3+k}} \big(\xi_1\xi_2\big)^{3-k}\big( \sum_{j=0}^{2+k} \xi_1^{j}{\xi_2}^{2+k-j}\big)\Big)&=&0\ ,\nonumber\\
\sum_{k=-3}^{2}\Big(\sum_{i_1<...<i_{3+k}}(-1)^{k}\chi_{i_1}\chi_{i_2}...\chi_{i_{3+k}}\big(\xi_1\xi_2\big)^{3+k}\big( \sum_{j=0}^{2-k} \xi_1^{j}{\xi_2}^{2-k-j}\big)\big)&=&0\ ,\label{prest}
\eeqa
with $\{i_k\}\in\{1,...,6\}$. So, it remains six independent parameters, say $\chi_1,\chi_2,\chi_3,\chi_4,\chi_5,\chi_6$. Simplifying (\ref{cond}), one finds
\beqa
\rho=-(q-q^{-1})^2\chi_1\chi_2\chi_3\chi_4\chi_5\chi_6\xi^{-1}_1\xi_2^{-1}q^{-1}\ .
\eeqa

Although we do not proceed further\,\footnote{The case $N=3$ has been also considered, although not reported here.}, for higher values of $N$ it is possible to classify all possible parameter sequences such that (\ref{real}) with (\ref{fonction}) and $d(z,z^{-1})\equiv Const.$ \ 
satisfy (\ref{qDG}). More generally, it is possible to find examples for $\phi,{\overline \phi}$ which are neither such that ${\overline \phi}(z)= \phi(z^{-1})$, nor {\it rational} functions of $z$.\vspace{1mm}

\subsection{A polynomial eigenbasis}
For the rational functions $\phi(z),{\overline \phi}(z)$ satisfying (\ref{cond}), a basis of possibly degenerate eigenfunctions $\psi_{n[m]}(z)$, $n=0,1,2,...$ and $m=1,...,dim(V_n)$ of ${\textsf W}_0$ with eigenvalues $\{\lambda_n\}$ can be constructed. The realization (\ref{real}) for ${\textsf W}_0$ yields to the $q-$Sturm-Liouville problem 
\beqa
\phi(z)\psi_{n[m]}(qz) + {\overline \phi}(z)\psi_{n[m]}(q^{-1}z) + \mu(z,z^{-1})\psi_{n[m]}(z) = \lambda_{n}\psi_{n[m]}(z)\ \label{qdiff0}
\eeqa
with
\beqa
\mu(z,z^{-1})= d(z,z^{-1})-\phi(z)-{\overline \phi}(z)\ .\nonumber
\eeqa

In general, only a part of the spectrum of $q-$difference operators can be algebraized, i.e. corresponds to polynomial eigenfunctions in $z$. However, according to the structure of the spectrum of ${\textsf W}_0,{\textsf W}_1$ for the finite dimensional case \cite{Ter01,Ter03}, we expect the operator ${\textsf W}_0$ in (\ref{real}) can be entirely algebraized  provided (\ref{cond}) are satisfied. For this reason, we are looking for polynomial eigenfunctions {\it symmetric} in the variable $z$ i.e. invariant under the replacement $z\rightarrow z^{-1}$:
\beqa
\psi_{n[m]}(z)=\prod_{j=1}^{n}(z-z_j)(z^{-1}-z_j) \ \qquad \mbox{for} \qquad z_j\in \{z_j^{(m)}\}\ ,  \label{psi}
\eeqa
with $m=1,...,dim(V_n)$, $n=0,1,2,...$ and $z_j$ denote the roots of the polynomials. Replacing this expression in (\ref{qdiff0}) and setting $z\rightarrow z_i$, an immediate consequence of the factorized structure (\ref{psi}) is the system of Bethe equations
\beqa
\frac{\psi_{n}(qz_i)}{\psi_{n}(q^{-1}z_i)}=-\frac{{\overline \phi}(z_i)}{\phi(z_i)}\qquad \mbox{with}\qquad i=1,2,...,n\label{BAprel}\ .
\eeqa 

As an example, let us consider the family of rational functions (\ref{fonction}) with suitable parameter sequences such that (\ref{cond}) are satisfied. Introducing the new parametrization
\beqa
z\equiv e^{2\lambda}\ ,\quad \chi_l\equiv e^{2\eta_l}\ ,\quad \xi_l\equiv e^{2c_l}\ ,\quad q\equiv e^{\varphi}\ ,\label{lien}
\eeqa
the Bethe equations take the form
\beqa
\prod_{l=1}^{2N-2}\frac{\sinh(\lambda_i+c_l)}{\sinh(\lambda_i-c_l)}\prod_{l=1}^{2N+2} \frac{\sinh(\lambda_i-\eta_l)}{\sinh(\lambda_i+\eta_l)}   =\prod_{j=1,j\neq i}^{n}\frac{\sinh(\lambda_i+\lambda_j+\varphi/2)\sinh(\lambda_i-\lambda_j+\varphi/2)}{\sinh(\lambda_i+\lambda_j-\varphi/2)\sinh(\lambda_i-\lambda_j-\varphi/2)}\ \label{BAfin}
\eeqa
with $i=1,2,...,n$. For $N=1$ and  no special relations among the parameters $\{\chi_k\}$, one has $dim(V_n)=1$. The polynomial eigenfunctions (\ref{psi}) coincide exactly with the Askey-Wilson polynomials which zeros are known to satisfy (\ref{BAfin}) at $N=1$. For $N=2$ the parameters $c_1,c_2$ are determined by the relations (\ref{prest}), and
the solutions (\ref{psi}) can be seen as some generalization of the Askey-Wilson polynomials which, to the best of our knowledge, are new. In both {\it exactly solvable} cases, the spectrum can be easily derived. Replacing (\ref{psi}) in (\ref{qdiff0}), expanding both sides of the equation and identifying the leading terms one finds
\beqa
\lambda_n&=& \chi_1\chi_2\chi_3\chi_4q^{-1} q^{n}+ q^{-n}\qquad \qquad \qquad  \qquad \ \ \ \ \quad \mbox{for}\qquad N=1\ ,\nonumber\\
\lambda_n&=& \chi_1\chi_2\chi_3\chi_4\chi_5\chi_6\xi_1^{-1}\xi_2^{-1}q^{-1}q^{n} + q^{-n}\qquad \ \ \ \qquad \mbox{for}\qquad N=2\ .
\eeqa 
For higher values of $N$, a spectrum of the form 
\beqa
{\textsf W}_0\ \psi_{n[m]}(z)&=&\lambda_n\ \psi_{n[m]}(z)\ \qquad \mbox{with} \qquad \lambda_n= C q^n+C'q^{-n} \ ,\label{var1}
\eeqa
$C=g(\{\chi_k\},\{\xi_k\},q)$, $C'=1$ is clearly expected: the eigenvalue sequences for $N=1,2$ and $N=3$ although not reported here agree with  (\ref{var1}). More generally, such form of the spectrum coincides exactly with the one associated with finite dimensional modules \cite{Ter01,Ter03} (see also \cite{TDpair}).\vspace{1mm} 

Let us now consider the action of ${\textsf W}_1$ on the eigenbasis of ${\textsf W}_0$. According to (\ref{real}) and the fact that $\psi_{n[m]}(z)$ are symmetric polynomials of degree $n$ in $x=z+z^{-1}$, the operator ${\textsf W}_1$ has a block tridiagonal structure in the eigenbasis of ${\textsf W}_0$. Its action on $\psi_{n[m]}(z)$ can be written formally\,\footnote{This property can be also derived starting from (\ref{qDG}). See for instance \cite{Ter03}.}
\beqa
{\textsf W}_1\psi_{n[m]}(z)&=& \sum_{l=1}^{dim(V_{n+1})}\bcal_{n[lm]}\psi_{n+1[l]}(z) + \sum_{l=1}^{dim(V_{n})}\acal_{n[lm]}\psi_{n[l]}(z) + \sum_{l=1}^{dim(V_{n-1})}\ccal_{n[lm]}\psi_{n-1[l]}(z)\ . \ \label{recpsi}
\eeqa

For the family of rational functions (\ref{fonction}) and $d(z,z^{-1})\equiv Const.$, the coefficients $\acal_{n[lm]},\bcal_{n[lm]},\ccal_{n[lm]}$ can be determined explicitly in terms of the parameters $\chi_k,\xi_k$. 
As the simplest example, consider the fundamental operators ${\textsf W}_0,{\textsf W}_1$ with (\ref{real}) and (\ref{fonction}) for $N=1$. To avoid degenerate situations, we assume the parameters $\{\chi_k\}$ are generic. As mentionned above, in this case $dim(V_n)=1$ i.e. $m,l=1$. The coefficients in (\ref{recpsi}) are well-known \cite{AW}, they take the form
\beqa
\bcal_{n[11]} &=& {{(1-\chi_1\chi_2q^n)(1-\chi_1\chi_3q^n)(1-\chi_1\chi_4q^n)(1-\chi_1\chi_2\chi_3\chi_4q^{n-1})}
\over {\chi_1(1-\chi_1\chi_2\chi_3\chi_4q^{2n-1})(1-\chi_1\chi_2\chi_3\chi_4q^{2n})}},\nonumber\\
\ccal_{n[11]} &=& {{\chi_1(1-q^n)(1-\chi_2\chi_3q^{n-1})(1-\chi_2\chi_4q^{n-1})(1-\chi_3\chi_4q^{n-1})}
\over {(1-\chi_1\chi_2\chi_3\chi_4q^{2n-2})(1-\chi_1\chi_2\chi_3\chi_4q^{2n-1})}},\nonumber\\
\acal_{n[11]} &=& \chi_1+\chi_1^{-1} -\bcal_{n[11]} -\ccal_{n[11]}\ .
\label{coeff}
\eeqa

To resume, polynomial eigenfunctions of the form (\ref{psi}) with Bethe equations (\ref{BAprel}) provide an example of irreducible\,\footnote{For special relations among the parameters entering in the solutions $\phi(z),{\overline \phi}(z),d(z,z^{-1})$, the module may become indecomposable. This possibility is not considered here.} infinite dimensional module of the tridiagonal algebra. In this basis, ${\textsf W}_0$, ${\textsf W}_1$ act as (\ref{var1}) and (\ref{recpsi}), respectively, where explicit expressions of the coefficients depend on the solutions $\phi(z),{\overline \phi}(z),d(z,z^{-1})$ of (\ref{cond}). For the family of rational functions (\ref{fonction}), the Bethe equations take the form (\ref{BAfin}) with the identifications (\ref{lien}).    

\subsection{Structure of the eigenfunctions}
For quantum integrable models with dynamical symmetry (\ref{qDG}), the spectral problem for the family of mutually commuting operators $\{{\cal I}_{2k+1}\}$ $k=0,1,...$, reduces to the one for (\ref{I1}) provided the spectrum of this operator is non-degenerate. Let $\Psi(z)$ denotes an eigenfunction of the hierarchy (\ref{IN}). Using the realization (\ref{real}) of the fundamental operators, the spectral problem for ${\cal I}_{1}$ in the functional representation reads\,\footnote{The reader familiar with the algebraic Bethe ansatz will immediatly recognize the similarity of this equation with the Baxter identity for the eigenvalues of the transfer matrix, also called T-Q relations \cite{Baxter}.}   
\beqa
A(z)\Psi(qz) + {\overline A}(z)\Psi(q^{-1}z) + B(z,z^{-1})\Psi(z) = \Lambda_{1}\Psi(z)\ \label{qdiff}
\eeqa
with
\beqa
A(z)&=&\Big(\kappa+(q-q^{-1})\big(\frac{\kappa_+}{k_+}q^{-1/2}z^{-1}+\frac{\kappa_-}{k_-}q^{1/2}z\big)\Big)\phi(z)\ ,\nonumber\\
{\overline A}(z)&=&\Big(\kappa+(q-q^{-1})\big(\frac{\kappa_+}{k_+}q^{-1/2}z+\frac{\kappa_-}{k_-}q^{1/2}z^{-1}\big)\Big){\overline \phi}(z)\ ,\nonumber\\
B(z,z^{-1})&=&\kappa\mu(z,z^{-1})+\kappa^*(z+z^{-1})+(q^{1/2}-q^{-1/2})(z+z^{-1})\Big(\frac{\kappa_+}{k_+}+\frac{\kappa_-}{k_-}\Big)\mu(z,z^{-1})\ .\nonumber
\eeqa
Here, the parameters $\kappa,\kappa^*,\kappa_\pm,k_\pm$ are arbitrary.
Contrary to the case of ${\textsf W}_0$, the operator ${\cal I}_1$ admits only {\it partial} algebraization of it spectrum. Depending on the parameters, we now describe the structure of its eigenfunctions $\Psi(z)$ in the non-algebraic sector and the algebraic sector, respectively. \vspace{1mm}

$\bullet$ {\bf Non-algebraic sector}: 
For generic values of the parameters $\kappa,\kappa^*,\kappa_\pm,k_\pm$, there are {\it no} polynomial solutions. Indeed, the leading contribution in the numerator of  (\ref{qdiff}) - independent of $\Lambda_1$ - can not be cancelled out. A natural approach is then to consider $\Psi(z)$ as an infinite power series expansion in the variable $x=z+z^{-1}$. However, for the family of rational functions (\ref{fonction}) the problem becomes quickly rather complicated when the value of $N$ increases. Instead, a convenient procedure consists in expanding the solutions of (\ref{qdiff}) on the basis of eigenfunctions $\psi_{n[m]}(z)$ of ${\textsf W}_0$ given by (\ref{psi}) with (\ref{BAfin}). For generic values of the parameter $\kappa,\kappa^*,\kappa_\pm,k_\pm$ we write
\beqa
\Psi(z)= \sum_{n=0}^{\infty}\sum_{m=1}^{dim(V_n)} f_{n[m]}(\Lambda_1)\psi_{n[m]}(z)\ .\label{Bethevec}
\eeqa
According to (\ref{real}) and the action (\ref{var1}) of ${\textsf W}_0$ and (\ref{recpsi}) of ${\textsf W}_1$ on the polynomial basis, the coefficients $f_{n[m]}(\Lambda_1)$ are determined by the following coupled system of recurrence relations:
\beqa
\big(\kappa\lambda_n-\Lambda_1\big)f_{n[l]}+ \sum_{m=1}^{dim(V_{n-1})}{\cal B}_{n-1[lm]}f_{n-1[m]} + \sum_{m=1}^{dim(V_n)}{\cal A}_{n[lm]}f_{n[m]} + \sum_{m=1}^{dim(V_{n+1})}{\cal C}_{n+1[lm]}f_{n+1[m]} =0\   ,\label{recf}
\eeqa
for $l=1,...,dim(V_n)$, where the coefficients ${\cal A}_{n[lm]},{\cal B}_{n[lm]},{\cal C}_{n[lm]}$ follow from the action of (\ref{I1}) on (\ref{psi}). Using (\ref{var1}) and (\ref{recpsi}), they read
\beqa
{\cal B}_{n[lm]}&=&\Big(\kappa^*+\big(\frac{\kappa_+}{k_+}q^{1/2}-\frac{\kappa_-}{k_-}q^{-1/2})\lambda_n + \big(\frac{\kappa_-}{k_-}q^{1/2}-\frac{\kappa_+}{k_+}q^{-1/2})\lambda_{n+1} \big)\Big)\bcal_{n[lm]}\ ,\nonumber\\ 
{\cal C}_{n[lm]}&=&\Big(\kappa^*+\big(\frac{\kappa_+}{k_+}q^{1/2}-\frac{\kappa_-}{k_-}q^{-1/2})\lambda_n + \big(\frac{\kappa_-}{k_-}q^{1/2}-\frac{\kappa_+}{k_+}q^{-1/2})\lambda_{n-1} \big)\Big)\ccal_{n[lm]}\ ,\nonumber\\ 
{\cal A}_{n[lm]}&=&\Big(\kappa^*+(q^{1/2}-q^{-1/2})\big(\frac{\kappa_+}{k_+}+\frac{\kappa_-}{k_-})\lambda_n\Big)\acal_{n[lm]}\ .\label{co}
\eeqa 

\vspace{1mm}

$\bullet$ {\bf Algebraic sector}: This sector corresponds to polynomial eigenfunctions of ${\cal I}_1$ which exist for special relations among the parameters, identified as follows. Let us assume that $\Psi_n(z)$ is a polynomial symmetric in the variable $z$ of the form    
\beqa
\Psi_{n}(z)=\prod_{j=1}^{n}(z-z_j)(z^{-1}-z_j) \ ,  \label{Psi}
\eeqa
with $n=0,1,...$. For the family of rational functions (\ref{fonction}) with suitable non-vanishing parameters, replacing (\ref{Psi}) in (\ref{qdiff}) one finds that the leading contribution in the l.h.s. of the equation (obtained by taking the limit $z\rightarrow\infty$) is independent of $\Lambda_1$. Then, the leading terms will cancel out (and similarly if one considers the limit $z\rightarrow 0$) only if the parameters satisfy the relations 
\beqa
\kappa^*+(q-q^{-1})\frac{\kappa_+}{k_+}q^{-1/2}q^{-n} + (q-q^{-1})\frac{\kappa_-}{k_-}q^{1/2}\Big(q^{-1}\prod_{k=1}^{2N+2}\chi_k\prod_{k=1}^{2N-2}\xi_k\Big)q^{n}=0\ ,\label{rel1}
\eeqa
i.e. ${\cal B}_{n[lm]}=0$. This implies that any eigenfunction (\ref{Psi}) can be written as a {\it finite} sum of elementary eigenfunctions (\ref{psi}). Choosing for simplicity the representation (\ref{Psi}), the roots $z_i$ are now determined by the generalized system of Bethe equations
\beqa
\frac{\Psi_{n}(qz_i)}{\Psi_{n}(q^{-1}z_i)}=-\frac{\Big(\kappa+(q-q^{-1})\big(\frac{\kappa_+}{k_+}q^{-1/2}z_i+\frac{\kappa_-}{k_-}q^{1/2}z_i^{-1}\big)\Big)}{\Big(\kappa+(q-q^{-1})\big(\frac{\kappa_+}{k_+}q^{-1/2}z_i^{-1}+\frac{\kappa_-}{k_-}q^{1/2}z_i\big)\Big)}\frac{{\overline \phi}(z_i)}{\phi(z_i)}\qquad \mbox{for}\qquad i=1,2,...,n\label{BAprel1}\ .
\eeqa 
Given $n$ fixed by (\ref{rel1}), the spectrum can be calculated by plugging (\ref{Psi}) into (\ref{qdiff}) and identifying the leading terms of both sides of the $q-$difference equation. For the family of rational functions (\ref{fonction}), the result is
\beqa
\Lambda_1&=& \big(\kappa{\cal F}_+^{(N)}-(q-q^{-1})\frac{\kappa_-}{k_-}q^{1/2}{\cal G}_+^{(N)}\big) q^{n}+\big(\kappa{\cal F}_-^{(N)}-(q-q^{-1})\frac{\kappa_+}{k_+}q^{-1/2}{\cal G}_-^{(N)}\big) q^{-n}\nonumber\\
&&+ \ (q-q^{-1})(q^{1/2}-q^{-1/2})\Big(\frac{\kappa_-}{k_-}{\cal F}_+^{(N)}q^{n}-\frac{\kappa_+}{k_+}{\cal F}_-^{(N)}q^{-n}\Big)\sum_{i=1}^{n}(z_i+z_i^{-1})\ ,\label{spectrum}
\eeqa 
where ${\cal F}_\pm^{(N)},{\cal G}_\pm^{(N)}$ are some functions of the parameters $\{\chi_k\},\{\xi_k\}$. For small values of $N$,
\beqa
{\cal F}_+^{(1)}&=&\chi_1\chi_2\chi_3\chi_4q^{-1}\ ,\quad {\cal F}_-^{(1)}=1\ , \quad {\cal G}_+^{(1)}=\sum_{i_1<i_2<i_3}^{4}\chi_{i_1}\chi_{i_2}\chi_{i_3}\ ,\quad {\cal G}_-^{(1)}=\sum_{i=1}^{4}\chi_{i}\ ,\nonumber\\ 
{\cal F}_+^{(2)}&=&\chi_1\chi_2\chi_3\chi_4\chi_5\chi_6\xi^{-1}_1\xi^{-1}_2 q^{-1}\ ,\quad {\cal F}_-^{(2)}=1\ ,\quad {\cal G}_-^{(2)}=\sum_{i=1}^{6}\chi_{i}-\xi_1-\xi_2\ ,\nonumber\\
{\cal G}_+^{(2)}&=&\Big(\sum_{i_1<i_2<i_3<i_4<i_5}^{4}\chi_{i_1}\chi_{i_2}\chi_{i_3}\chi_{i_4}\chi_{i_5}\Big)\xi^{-1}_1\xi^{-1}_2 -\chi_1\chi_2\chi_3\chi_4\chi_5\chi_6\xi^{-1}_1\xi^{-1}_2 q^{-1}(\xi^{-1}_1+\xi^{-1}_2)\ .\nonumber
\eeqa
For higher values of $N$, a similar form of the spectrum is expected according to (\ref{fonction}). It is however important to remind that the parameters $\{\xi_k\}$ are restricted by some relations generalizing  (\ref{prest}). For non-vanishing parameters $\kappa_\pm$, the spectrum (\ref{spectrum}) is non-degenerate. So, (\ref{Psi}) is also a polynomial eigenfunction of the infinite hierarchy (\ref{IN}).\vspace{1mm}
     
Consequently, the operator ${\cal I}_1$ admits partial algebraization of its spectrum. In the algebraic sector associated with (\ref{rel1}), there is a single eigenstate associated with the polynomial eigenfunction (\ref{Psi}) with (\ref{BAprel1}) . For the family of rational functions (\ref{fonction}), the spectrum can be written in terms of the roots of the Bethe equations, and takes the form (\ref{spectrum}). The rest of the spectrum is associated with non-polynomial eigenfunctions which can be written in the (infinite dimensional) eigenbasis (\ref{psi}) with (\ref{BAprel}) of ${\textsf W}_0$. The structure of the eigenstates (\ref{Bethevec}) is encoded in the coefficients $f_{n[m]}$ which satisfy (\ref{recf}).

\section{A discrete $q-$difference equation for the XXZ open spin chain}
For quantum integrable lattice models with dynamical $q-$Onsager symmetry, a similar analysis can be done in order to derive a $q-$difference equation 
for the eigenstates of the transfer matrix (\ref{tfin}). The main difference is however the Hilbert space of these models, which is now finite dimensional\,\footnote{Linear relations among the higher operators of the $q-$deformed analogue of the Onsager algebra - generated from ${\textsf W}_0,{\textsf W}_1$ - exist that are responsible of the truncation of the hierarchy (\ref{IN}) (see \cite{qOns} for details).}. 
To give an illustration, let us consider the XXZ open spin$-\frac{1}{2}$ chain with general integrable boundary conditions. Its Hamiltonian reads 
\beqa
H_{XXZ}&=&\sum_{k=1}^{N-1}\Big(\sigma_1^{k}\sigma_1^{k+1}+\sigma_2^{k}\sigma_2^{k+1} + \Delta\sigma_3^{k}\sigma_3^{k+1}\Big) \nonumber\\
&&+\ \frac{(q^{1/2}-q^{-1/2})}{(\epsilon^{(0)}_+ + \epsilon^{(0)}_-)}
\Big(  \frac{(\epsilon^{(0)}_+ - \epsilon^{(0)}_-)}{2}\sigma^1_3 + \frac{2}{(q^{1/2}-q^{-1/2})}\big(k_+\sigma^1_+ + k_-\sigma^1_-\big)       \Big)\nonumber\\
 &&+\ \frac{(q^{1/2}-q^{-1/2})}{(\kappa + \kappa^*)}
\Big(  \frac{(\kappa - \kappa^*)}{2}\sigma^N_3 + 2(q^{1/2}+q^{-1/2})\big(\kappa_+\sigma^N_+ + \kappa_-\sigma^N_-\big)       \Big)\label{H}\ ,
\eeqa
where $\sigma_{1,2,3}$ and $\sigma_\pm=(\sigma_1\pm i\sigma_2)/2$ are usual Pauli matrices. Here, $\Delta=(q^{1/2}+q^{-1/2})/2$\ \ denotes the anisotropy parameter with $q=\exp\varphi$. Integrable boundary conditions correspond to the choice \cite{DeVeg}
\beqa
&&\epsilon^{(0)}_{+}=(\epsilon^{(0)}_{-})^{\dagger}=(c_{00}+ic_{01})/2\ ,\qquad k_+=(k_{-})^{\dagger}=-(q^{1/2}-q^{-1/2})e^{i\theta}/2 \qquad\ \  \mbox{(left)}\ ,\nonumber\\ 
&&\kappa=(\kappa^*)^{\dagger}=(-{\tilde{c}}_{00}+i{\tilde{c}}_{01})/2\ ,\qquad
\kappa_+=(\kappa_{-})^{\dagger}=-e^{i{\tilde{\theta}}}/(2(q^{1/2}+q^{-1/2}))\qquad \mbox{(right)}\ .\label{param}
\eeqa
The r.h.s parametrization has been introduced in which case
one immediatly identifies the Hamiltonian as defined in \cite{Cao}. In total, one has {\it six} boundary parameters $c_{00},c_{01},{\tilde{c}}_{00},{\tilde{c}}_{01},\theta,{\tilde\theta}$. Except if explicitely specified, from now on we assume these parameters are {\it generic}.\vspace{1mm}

As discovered in \cite{TriDiag,qOns}, the known integrability of the XXZ open spin chain (\ref{H}) with (\ref{param}) \cite{Skly88,DeVeg} is related with 
the existence of a hidden {\it dynamical} symmetry, a $q-$deformed analogue of the Onsager algebra. This non-Abelian algebra is generated by a family of nonlocal operators $\{{\cal W}^{(N)}_{-k},{\cal W}^{(N)}_{k+1},{\cal G}^{(N)}_{k+1},{\tilde{\cal G}}^{(N)}_{k+1}\}$, which fundamental ones satisfy the tridiagonal relations (\ref{qDG}). They read \cite{qOns}
\beqa
{\cal W}^{(N)}_0&=& (k_+\sigma_+ + k_-\sigma_-)\otimes I\!\!I^{(N-1)} + q^{\sigma_3/2}\otimes {\cal W}_0^{(N-1)}\ ,\nonumber \\
{\cal W}^{(N)}_1&=& (k_+\sigma_+ + k_-\sigma_-)\otimes I\!\!I^{(N-1)} + q^{-\sigma_3/2}\otimes {\cal W}_1^{(N-1)}\ ,\label{op}
\eeqa
with ``initial'' conditions 
\beqa
{\cal W}^{(1)}_0&=& k_+\sigma_+ + k_-\sigma_- +  \epsilon_+^{(0)} q^{\sigma_3/2}\ ,\nonumber \\
{\cal W}^{(1)}_1&=& k_+\sigma_+ + k_-\sigma_- +  \epsilon_-^{(0)}q^{-\sigma_3/2}\ .\nonumber
\eeqa
Note that the left boundary parameters in (\ref{H}) appear in the realization (\ref{op}) of the algebra (\ref{qDG}), whereas the right boundary parameters enter in the definition of the linear combinations (\ref{IN}). For the normalization above, the structure constant of the tridiagonal algebra (\ref{qDG}) is fixed to
\beqa
\rho=\rho^*=-\frac{(q-q^{-1})^2}{4}\ .
\eeqa

Irreducible finite dimensional modules $V$ of the tridiagonal algebra (\ref{qDG}) associated with ${\cal W}^{(N)}_0,{\cal W}^{(N)}_1$ have been studied in details in \cite{TDpair}. Here, we just recall some results. Suppose the boundary parameters entering in (\ref{op}) are such that both operators are diagonalizable on $V$, and suppose $q\equiv e^\varphi$ - with $\varphi$ purely imaginary - is not a root of unity.  
Then, according to [\cite{Ter03}, Theorem 3.10] the operators ${\cal W}^{(N)}_0,{\cal W}^{(N)}_1$  act on $V$ as a tridiagonal pair: The pair of linear transformations ${\cal W}^{(N)}_0: \ V \rightarrow V$ and ${\cal W}^{(N)}_1:\ V \rightarrow V$ satisfy the following \vspace{1mm}

$(i)$ There exists an ordering $V_0, V_1,\ldots, V_N$ of the  
eigenspaces of ${\cal W}^{(N)}_0$ such that 
\beqa
{\cal W}^{(N)}_1 V_n \subseteq V_{n+1} + V_n+ V_{n-1} \qquad \qquad (0 \leq n \leq N),\qquad \mbox{and} \qquad V_{-1} = 0, \ V_{N+1}= 0\ ,\nonumber
\eeqa

$(ii)$ There exists an ordering $V^*_0, V^*_1,\ldots, V^*_N$ of
the  
eigenspaces of ${\cal W}^{(N)}_1$ such that 
\beqa
{\cal W}^{(N)}_0 V^*_s \subseteq V^*_{s+1} + V^*_s+ V^*_{s-1} \qquad \qquad (0 \leq s \leq N),\qquad \mbox{and} \qquad V^*_{-1} = 0, \ V^*_{N+1}= 0\ .\nonumber
\eeqa
Also, for $0 \leq n \leq N$, the dimensions of the eigenspaces $V_n$ and $V^*_s$ are equal and the {\it shape vector} \cite{Ter01} of the pair is symmetric and unimodal. For the family of operators (\ref{op}), one has\,\footnote{We denote $\left({{ N }\atop {n}}\right)=\frac{N!}{n!(N-n)!}$\ .} \cite{TDpair} 
\beqa
dim(V_n) = \biggl({{N}\atop {n}}\biggr), \qquad 0 \leq n \leq N\ .\label{Vn}
\eeqa

Depending on the basis chosen the matrix representing ${\cal W}^{(N)}_0$ (resp. ${\cal W}^{(N)}_1$) is either diagonal (with possible degeneracies) or (block) tridiagonal in the dual basis. Below, we will use these properties and refer the reader to the work \cite{TDpair} for the explicit expressions of the matrix representation in each basis. 

\subsection{Functional representation}
For our purpose, it will be useful to consider the functional representation of dimension $dim(V)=2^N$ defined on a discrete support associated with the overlap coefficients between the two dual eigenbasis of  ${\cal W}^{(N)}_0$, ${\cal W}^{(N)}_1$. As follows from $(i)$ and $(ii)$, these coefficients satisfy a set of coupled recurrence relations and a set of discrete $q-$difference equations, respectively. Introducing the parametrization
\beqa
\epsilon_+^{(0)}=\cosh\alpha \qquad \mbox{and}\qquad \epsilon_-^{(0)}=\cosh\alpha^*\ ,\label{defb2}
\eeqa 
they are denoted 
\beqa
\psi^{(N)}_{n[m]}(z_s) \qquad \mbox{with} \qquad z_s=\exp(\alpha^*+(N-2s)\varphi/2)\ ,\label{frep}
\eeqa
where $n=0,1,...,N$ and $m=1,...,\bigl({{N}\atop {n}}\bigr)$ characterizes the degeneracy of the eigenvalue with index $n$. Note that these coefficients are symmetric in the variable $z_s$, as we will see below. 
They provide a functional eigenbasis for ${\cal W}^{(N)}_0$. Indeed, one has \cite{TDpair} 
\beqa
{\cal W}^{(N)}_0\psi^{(N)}_{n[m]}(z_s)&=&\lambda_n\psi^{(N)}_{n[m]}(z_s) \qquad \mbox{with} \qquad \lambda_n=\cosh(\alpha+(N-2n)\varphi/2)\ .\label{spec0}
\eeqa

On the other hand, the action of the operator ${\cal W}^{(N)}_1$ in the basis (\ref{frep}) can be explicitely calculated. It takes a block tridiagonal structure which is formally written as  
\beqa
{\cal W}^{(N)}_1\psi^{(N)}_{n[m]}(z_s)&=& \sum_{l=1}^{\bigl({{N}\atop {n+1}}\bigr)}\bcal_{n[lm]}\psi^{(N)}_{n+1[l]}(z_s) + \sum_{l=1}^{\bigl({{N}\atop {n}}\bigr)}\acal_{n[lm]}\psi^{(N)}_{n[l]}(z_s) + \sum_{l=1}^{\bigl({{N}\atop {n-1}}\bigr)}\ccal_{n[lm]}\psi^{(N)}_{n-1[l]}(z_s)\ , \ \label{recpsidisc}
\eeqa
where the explicit expressions of the matrix entries can be found in \cite{TDpair}. Analogous results also hold if, instead, we had considered the dual overlap coefficients which also form a complete eigenbasis of ${\cal W}^{(N)}_1$, a phenomena due to the duality property of (\ref{qDG}).  Obviously, in this dual basis the operator ${\cal W}^{(N)}_0$ has a block tridiagonal structure.\vspace{1mm}

The analysis of previous Section can be applied to the finite dimensional case similarly: the operator ${\cal W}^{(N)}_0$ can be associated with a discrete $q-$difference operator acting on the (discrete) functional space spanned by (\ref{frep}). Defining $\eta^{\pm1}(f(z_s))\equiv f(z_{s\pm 1})=f(q^{\mp 1}z_s)$, the discrete analogue of the realization (\ref{real}) is now given by \cite{TDpair}
\beqa
{\cal W}^{(N)}_0:&& \qquad \Phi_k^{(N)}(s)\eta_s + \overline{\Phi}_k^{(N)}(s)\eta_s^{-1} + \mu_k^{(N)}(s)\ ,\qquad k\in \{1,...,\Big({{N}\atop {s}}\Big)\}\ ,\nonumber\\
{\cal W}^{(N)}_1:&& \qquad (z_s + z_s^{-1})/2\ ,\label{qdiffop}
\eeqa
where 
\beqa 
\Phi_k^{(N)}(s)&=&\sum_{l=1}^{({{N}\atop {s+1}})}\bcalt^{(N)}_{s[lk]}\frac{{ U}^{(N)}_l(s+1)}{{ U}^{(N)}_k(s)}\ ,\qquad \overline{\Phi}_k^{(N)}(s)=\sum_{l=1}^{({{N}\atop {s-1}})}\ccalt^{(N)}_{s[lk]}\frac{{ U}^{(N)}_l(s-1)}{{ U}^{(N)}_k(s)}\ ,\nonumber\\
\mu_k^{(N)}(s)&=&\sum_{l=1}^{({{N}\atop {s}})}\acalt^{(N)}_{s[lk]}\frac{{ U}^{(N)}_l(s)}{{ U}^{(N)}_k(s)}\ .\nonumber
\eeqa
The explicit expressions of the coefficients can be found in \cite{TDpair}. It is important to notice that the realization of ${\cal W}^{(N)}_0$ is not unique (except for $s=0$ and $s=N$), as the spectrum of ${\cal W}^{(N)}_1$ contains $\Big({{N}\atop {s}}\Big)$ degeneracies. 

\subsection{Eigenstates of the XXZ open spin chain}
Due to the non-Abelian $q-$Onsager dynamical symmetry of XXZ open spin chain (\ref{H}), the transfer matrix can be written in the form (\ref{tfin}) with (\ref{IN}) \cite{TriDiag,qOns}. By analogy with the case of infinite dimensional modules, all necessary and sufficient information to determine the eigenstates of (\ref{tfin}) is encoded in the structure of the tridiagonal pair ${\cal W}^{(N)}_0,{\cal W}^{(N)}_1$ given by (\ref{op}). We choose to take (\ref{frep}) as the basis on which any eigenstate, denoted $\Psi(z_s)$, is expanded. 
Considering the spectral problem for ${\cal I}_1$ and using the realization (\ref{qdiffop}), we find that the eigenstates of the XXZ open spin chain with general boundary conditions (\ref{H}) are the family of solutions to the following system of discrete $q-$difference equations 
\beqa
A_k^{(N)}(z_s)\Psi(qz_s) + \overline{A}_k^{(N)}(z_s)\Psi(q^{-1}z_s) + B_k^{(N)}(z_s,z_s^{-1})\Psi(z_s) =\Lambda_{1}\Psi(z_s)\ ,\qquad k=1,...,\Big({{N}\atop {s}}\Big)\label{qdisc}
\eeqa 
with
\beqa
A_k^{(N)}(z_s)&=&\Big(\kappa+(q-q^{-1})\big(\frac{\kappa_+}{2k_+}q^{-1/2}z_s^{-1}+\frac{\kappa_-}{2k_-}q^{1/2}z_s\big)\Big)\overline{\Phi}_k^{(N)}(s)\ ,\nonumber\\
{\overline A}_k^{(N)}(z_s)&=&\Big(\kappa+(q-q^{-1})\big(\frac{\kappa_+}{2k_+}q^{-1/2}z_s+\frac{\kappa_-}{2k_-}q^{1/2}z_s^{-1}\big)\Big)\Phi_k^{(N)}(s)\ ,\nonumber\\
B_k^{(N)}(z_s,z_s^{-1})&=&\kappa\mu_k^{(N)}(s)+\Big(\kappa^*+(q^{1/2}-q^{-1/2})\big(\frac{\kappa_+}{k_+}+\frac{\kappa_-}{k-}\big)\mu_k^{(N)}(s)\Big)\frac{(z_s+z_s^{-1})}{2}\ ,\nonumber
\eeqa
(\ref{qdiffop}) and $s=0,1,...,N$. \vspace{1mm}

Difference equations are known to be closely related with discrete equations. For the example considered here, the $q-$difference equation\,\footnote{The class of $q-$difference equations (\ref{qdiff}) may be extended by a ``gauge'' transformation $\Psi(z)\rightarrow f(z)\Psi(z)$, $\phi(z)\rightarrow \phi(z)f(z)/f(qz)$, $\phi(z)\rightarrow {\overline\phi}(z)f(z)/f(q^{-1}z)$ and $\mu(z)\rightarrow \mu(z)$, where $f(z)$ is a rational function of $z$.} (\ref{qdiff}) can be seen as an extension of (\ref{qdisc}) on the whole complex plane, where the parameters $\{\chi_k\},\{\xi_k\}$ in (\ref{fonction})    are chosen such that
\beqa
\phi(z=z^{-1}_s)\rightarrow \Phi_k^{(N)}(s)\ ,\quad {\overline\phi}(z=z^{-1}_s)\rightarrow {\overline\Phi}_k^{(N)}(s)\ ,\quad  {\mu}(z,z^{-1})|_{z=z^{-1}_s}\rightarrow {\mu}_k^{(N)}(s)\ ,\qquad \Psi(z=z^{-1}_s)\rightarrow \Psi(z_s)\ .\label{corres}
\eeqa
Indeed, finite dimensional module of the tridiagonal algebra (\ref{qDG}) can be obtained from the infinite dimensional modules by imposing $z$ to be on a discrete support $z=z_s$, $s=0,1,...,N$.  It follows that the results of previous Section can be used to analyse the structure of the eigenstates of the model (\ref{H}). In particular, in view of (\ref{rel1}), one expects that these eigenstates admit a description in terms of solutions of Bethe equations for certain relations between the boundary parameters. Below, we call the corresponding special cases the algebraic sector of the operator (\ref{I1}).\vspace{2mm}   

$\bullet$ {\bf Non-algebraic sector}: For generic values of the boundary parameters $\kappa,\kappa^*,\kappa_\pm,k_\pm$, there is {\it no} non-trivial subspace of $V$ which is left invariant by the action of (\ref{I1}). So, any eigenstate is a linear combination of {\it all} the eigenfunctions (\ref{frep}), $n=0,1,...,N$. In general, it takes the form
\beqa
\Psi(z_s)= \sum_{n=0}^{N}\sum_{m=1}^{\big({{N}\atop {n}}\big)} f_{n[m]}(\Lambda_1)\psi^{(N)}_{n[m]}(z_s)\ .\label{BethevecNdisc}
\eeqa
The eigenfunctions (\ref{frep}) being clearly identified \cite{TDpair}, it remains to determine the non-vanishing coefficients $f_{n[m]}$. Using the block tridiagonal structure (\ref{recpsidisc}) of ${\cal W}^{(N)}_1$ in the functional basis (\ref{frep}), we find that the coefficients $f_{n[m]}(\Lambda_1)$ satisfy the set of coupled recurrence relations (\ref{recf}) with the definitions (\ref{spec0}), (\ref{Vn}) and the explicit expression for the coefficients taken from  \cite{TDpair}. Note that each eigenfunction (\ref{psi}) at $z=z_s$ is individually associated with a system of Bethe equations, whereas there is no single system of Bethe equations for $\Psi(z_s)$.\vspace{1mm} 

For small values of $N$, it is not difficult to check the recurrence relations (\ref{recf}) numerically\,\footnote{K. Koizumi, private communication.}. One finds that the $2^N$ values of $\Lambda_{1}$ agree exactly with the direct diagonalization of ${\cal I}_1$ for generic values of the boundary parameters (\ref{param}). As expected, the corresponding eigenvectors diagonalize the higher operators (\ref{IN}).\vspace{2mm} 
 
$\bullet$ {\bf Algebraic sectors}: The advantage of the formulation of the transfer matrix in terms of (\ref{IN}) is that the representation theory of the tridiagonal algebra (\ref{qDG}) guarantees that the discrete $q-$difference equations (\ref{qdisc}) admit a subset of solutions associated with Bethe equations. This subset is identified with the special points in (\ref{qdiff}) with (\ref{spec0}), (\ref{Vn}) - or equivalently (\ref{qdisc}) using (\ref{corres}) - where either some of the coefficients $A_k^{(N)}(z_s),{\overline A}_k^{(N)}(z_s)$ or ${\cal B}_{n[m]}$, ${\cal C}_{n[m]}$ in (\ref{co}) are vanishing, or the special points for which the eigenfunctions (\ref{frep}) are no longer linearly independent.\vspace{1mm} 

For instance, suppose ${\cal B}_{n[lm]}\equiv 0$. Plugging the parametrization (\ref{param}), (\ref{defb2}), (\ref{spec0}) into (\ref{co}),  one obtains a linear relation between the left and right boundary parameters which reads
\beqa
\alpha+{\tilde\alpha} = i({\tilde \theta}-\theta) - (N-2n+1)\varphi/2 \qquad mod(2i\pi)\ .\label{r1}
\eeqa
According to the action of the operator ${\cal I}_1$ on the eigenfunctions of ${\cal W}_0^{(N)}$:
\beqa
{\cal I}_1\psi^{(N)}_{p[m]}(z_s)= \kappa\lambda_p\psi_{p[m]}(z_s)+\sum_{l=1}^{\bigl({{N}\atop {p+1}}\bigr)}{\cal B}_{p[lm]}\psi^{(N)}_{p+1[l]}(z_s) + \sum_{l=1}^{\bigl({{N}\atop {p}}\bigr)}{\cal A}_{p[lm]}\psi^{(N)}_{p[l]}(z_s) + \sum_{l=1}^{\bigl({{N}\atop {p-1}}\bigr)}{\cal C}_{p[lm]}\psi^{(N)}_{p-1[l]}(z_s)\ , \ \label{recI1}
\eeqa
one has
\beqa
{\cal I}_1 V_n \subseteq V_{n}+V_{n-1}\ ,\quad
{\cal I}_1 V_{n-1} \subseteq V_{n}+V_{n-1}+V_{n-2}\ ,\quad ...\quad \ , \quad {\cal I}_1 V_0 \subseteq V_{1}+V_{0}\ .\nonumber
\eeqa
Then, 
\beqa
{\cal I}_1W_n\subseteq W_n \qquad \mbox{where}\qquad W_n=\bigoplus_{p=0}^{n}V_p\nonumber
\eeqa
is an invariant eigenspace of (\ref{IN}).  The eigenstates $\Psi_n(z_s)$ associated with (\ref{Psi}) at $z=z_s$ and (\ref{BAprel1}) form an eigenbasis of $W_n$ which dimension follows from the degeneracies of each eigenspace of ${\cal W}_0^{(N)}$, given by (\ref{Vn}). One has 
\beqa
dim(W_n)=\sum_{p=0}^{n}\biggl({{N}\atop {p}}\biggr)\ .\label{Wn} 
\eeqa

This invariant subspace of (\ref{IN}) is not unique: another invariant subspace 
\beqa
{\tilde W}_n=\bigoplus_{p=n}^{N}V_p\qquad \mbox{with}\qquad dim({\tilde W}_n)=\sum_{p=n}^{N}\biggl({{N}\atop {p}}\biggr)\ \label{Wnautre}\nonumber
\eeqa
exists provided the boundary parameters are suitably tuned such that ${\cal C}_{n[lm]}\equiv 0$. Using (\ref{param}), (\ref{defb2}), (\ref{spec0}) in (\ref{co}), this condition corresponds to
\beqa
\alpha+{\tilde\alpha} = i(\theta-{\tilde \theta}) - (N-2n-1)\varphi/2 \qquad mod(2i\pi)\ .\label{r2}
\eeqa
In this case, using the substitution $n\rightarrow N-n$ in (\ref{Psi}) at $z=z_s$, the eigenstates with  (\ref{BAprel1}) form a basis of ${\tilde W}_n$.
It is important to notice that the conditions \ $A_k^{(N)}(z_s)\equiv0$ \ or \ ${\overline A}_k^{(N)}(z_s)\equiv0$ \ may have been considered alternatively, in which case the linear relations between the boundary parameters are obtained from (\ref{r1}), (\ref{r2}) by complex conjugation. This is not surprising, in view of the duality between the eigenbasis of (\ref{op}). To complete the analysis of the eigenstates in both algebraic sectors (\ref{r1}), (\ref{r2}), it is worth mentionning that the solutions $\Psi_n(z_s)$ can be regarded as linear combinations of the eigenfunctions (\ref{frep}), truncated in comparison with (\ref{BethevecNdisc}). The coefficients in the expansion satisfy a set of recursion relations, reported in Appendix A.\vspace{1mm}

Other invariant subspaces exist if the eigenfunctions (\ref{frep}) entering in (\ref{recI1}) are not linearly independent. These are the special points where the spectrum (\ref{spec0}) - or the one in the dual basis - admits additional degeneracies, namely
\beqa
\alpha + (N-2n)\varphi/2=i\pi {\mathbb Z}\ ,\label{r3}
\eeqa
in perfect agreement with \cite{Rittenberg}. The same result can be also derived from the explicit form of the eigenfunctions $\psi_{n[m]}(z_s)$ \cite{TDpair}.\vspace{1mm}

To conclude this Section, the approach based on the tridiagonal algebra (\ref{qDG}) provides a new derivation of the linear relations (\ref{r1}), (\ref{r2}), (\ref{r3}). In our framework, the (Bethe) eigenstates $\Psi_n(z_s)$ correspond to the algebraic sectors of the $q-$deformed Dolan-Grady hierarchy (\ref{IN}), where 
the Bethe anstaz equations have been formulated and the spectrum of the XXZ open spin chain has been derived \cite{Nepo1,Cao,Nepo2}.

\section{Comments}
A new approach based on the algebraic properties and irreducible modules of the infinite dimensional $q-$deformed analogue of the Onsager algebra has been proposed in the context of quantum integrable models. The main result of this paper is an exact second-order $q-$difference equation (resp. its degenerate discrete version) for the eigenfunctions associated with the infinite (resp. truncated) $q-$deformed Dolan-Grady hierarchy (\ref{IN}). For generic values of the parameters $\kappa,\kappa^*,\kappa_\pm,k_\pm$ and variable $z$ (resp. discrete $z_s$), the structure of the eigenfunctions has been described in details. They admit an infinite (resp. finite) serie expansion in terms of the polynomial eigenfunctions of ${\textsf W}_0$ which roots satisfy a system of Bethe equations of the form (\ref{BAfin}), whereas the weight functions satisfy a coupled system of recurrence relations which explicit form is determined from ${\textsf W}_1$. In the algebraic sectors which correspond to special relations between the parameters, the expansion truncates. In this case, the eigenstates admit a formulation in terms of solutions of the system of Bethe equations (\ref{BAprel1}).
The approach presented here has been applied to the XXZ open spin chain, although higher spin representations can be obviously studied similarly.\vspace{1mm}

Equations of the type (\ref{qdiff}), (\ref{qdisc}) already have important physical applications. For instance, the method of algebraization has been applied to the Azbel-Hofstader problem \cite{Wieg} (see also \cite{Wieg2}). In this model, the spectral problem for the Hamiltonian (Schr${\ddot{o}}$dinger equation) $H\equiv{\cal I}_1$ with anisotropy parameters $\kappa,\kappa^*$  is the famous Harper's equation. For certain values of the physical parameters, it can be algebraized using its connection with $U_{q}(sl_2)$ at root of unity. Non-polynomial solutions have been considered in \cite{Micu}.\vspace{1mm}
  
Such a connection between $q-$Sturm-Liouville problems and quantum integrable systems is not new. However, its understanding and formulation using the $q-$deformed Onsager algebra/tridiagonal algebra (\ref{qDG}) studied here is rather promising. A classical counterpart of our description clearly needs further investigations. In this direction, let us mention that a classical analogue of (\ref{qDG}) has been studied in \cite{Zhedclass}. The simplest example of realization of classical {\it Leonard pairs} ${\textsf W}_0$, ${\textsf W}_1$ has been proposed:
\beqa
{\textsf W}_0= y^2 + U(x) \qquad \mbox{and} \qquad {\textsf W}_1=\phi(x)\ \qquad \mbox{with}\qquad \{x,y\}=1\ ,
\eeqa 
where the admitted potentials $U(x)$ are the $(i)$ hyperbolic, $(ii)$ modified hyperbolic $(iii)$ trigonometric P${\ddot{o}}$schl-Teller potential, or the  $(iv)$ Morse, $(v)$ singular and $(vi)$ shifted oscillator potential. Indeed, in the continuum limit $q\rightarrow 1$ of (\ref{real}) with (\ref{fonction}) at $N=1$ the $q-$difference equation (\ref{qdiff}) turns into the  Schr${\ddot{o}}$dinger equation for the generalized P${\ddot{o}}$schl-Teller potential. Furthermore, using a suitable change of variable and ``gauge'' transformation one arrives at the Heun equation. In view of the recent works \cite{BLZ}, we expect our approach will provide an algebraic understanding of the ordinary-differential equations/integrable models (ODE/IM) (for a review, see \cite{ODEIM} and references therein).  
In the quantum case, it should be also mentionned that our results exhibit some links with a recent work relating a $q-$Sturm-Liouville problem with the Bethe ansatz equations of the XXZ open spin chain with {\it Dirichlet} boundary conditions \cite{Ro05}. In particular, for vanishing parameters $\{\xi_k\}$ the rational functions (\ref{fonction}) considered here can be easily related with the ones in \cite{Ismail}. However, for this choice the $q-$Onsager dynamical symmetry is broken - as expected - and (\ref{real}) do no longer satisfy (\ref{qDG}).\vspace{1mm}

To conclude, the program initiated in \cite{qDG} opens the possibility of studying massive quantum integrable models from an algebraic point of view (model-independent) which idea takes its roots in the original work of Onsager \cite{Ons}. The exact spectrum of the complete hierarchy (\ref{IN}) is the subject of a separate work, among other problems mentionned above.

\vspace{3mm}

{\underline {\bf Note added:}}
The explicit expressions for the entries of the matrices (\ref{op}) in their dual eigenbasis are not reported here. We refer the reader to \cite{TDpair} for these data.

\vspace{3mm}

\noindent{\bf Acknowledgements:} I thank  P. Forgacs, H. Giacomini, N. Kitanine, K. Koizumi and R. Weston for discussions, and P. Terwilliger for suggestions and interest in this work. I wish to thank the organizers of the 3rd Annual EUCLID Meeting 2005 where preliminary results were presented. Part of this work is supported by the ANR research project ``{\it Boundary integrable models: algebraic structures and correlation functions}'', contract number JC05-52749 and TMR Network EUCLID ``{\it Integrable models and applications: from strings to condensed matter}'', contract number HPRN-CT-2002-00325.\vspace{0.5cm}

\vspace{1cm}

\centerline{\bf \large Appendix A: An alternative construction of the Bethe eigenstates}
\vspace{3mm}

For special relations between the left and right boundary parameters, the corresponding Bethe eigenstates can be written directly in terms of the eigenfunctions of ${\cal W}_0^{(N)}$.
Consider the linear combination 
\beqa
\Psi_n(z_s)= \sum_{p=0}^{n}\sum_{m=1}^{\big({{N}\atop {p}}\big)} f'_{p[m]}(\Lambda_1)\psi^{(N)}_{p[m]}(z_s) \qquad \mbox{for}\qquad n\leq N-1 \ ,\label{tpsi1}
\eeqa
where $f'_{p[m]}$ are non-vanishing coefficients. As an immediate consequence of (\ref{recI1}), given $n$ this combination is an eigenfunction of ${\cal I}_1$ iff the condition ${\cal B}_{n[lm]}\equiv 0$ i.e. (\ref{r1}) and the coefficients for $p=0,1,...,n-1$ with degeneracies $l=1,...,\big({{N}\atop {p}}\big)$ satisfy
\beqa
\big(\kappa\lambda_p-\Lambda_1\big)f'_{p[l]}+ \sum_{m=1}^{\big({{N}\atop {p-1}}\big)}{\cal B}_{p-1[lm]}f'_{p-1[m]} + \sum_{m=1}^{\big({{N}\atop {p}}\big)}{\cal A}_{p[lm]}f'_{p[m]} + \sum_{m=1}^{\big({{N}\atop {p+1}}\big)}{\cal C}_{p+1[lm]}f'_{p+1[m]} =0\   ,\label{recfprime0}
\eeqa
and 
\beqa
\big(\kappa\lambda_n-\Lambda_1\big)f'_{n[l]}+ \sum_{m=1}^{\big({{N}\atop {n}}\big)}           {\cal A}_{n[lm]}f'_{n[m]} + \sum_{m=1}^{\big({{N}\atop {n-1}}\big)}{\cal C}_{n+1[lm]}f'_{n+1[m]} =0\  \qquad \mbox{for}\qquad l=1,...,\Big({{N}\atop {n}}\Big)\ .\label{recfprime1}
\eeqa

Antoher choice is ${\cal C}_{n[lm]}\equiv 0$, which corresponds to the linear relation between the left and right boundary parameters (\ref{r2}).
An eigenfunction of ${\cal I}_1$ takes the form
\beqa
\Psi_n(z_s)= \sum_{p=n}^{N}\sum_{m=1}^{\big({{N}\atop {p}}\big)} f'_{p[m]}(\Lambda_1)\psi^{(N)}_{p[m]}(z_s) \qquad \mbox{for}\qquad n\geq 1 \ ,\label{tpsi2}
\eeqa
where the coefficients in the expansion for $p=n+1,...,N$ with degeneracies $l=1,...,\big({{N}\atop {p}}\big)$ are determined by the set of recurrence relations
\beqa
\big(\kappa\lambda_p-\Lambda_1\big)f'_{p[l]}+ \sum_{m=1}^{\big({{N}\atop {p-1}}\big)}{\cal B}_{p-1[lm]}f'_{p-1[m]} + \sum_{m=1}^{\big({{N}\atop {p}}\big)}{\cal A}_{p[lm]}f'_{p[m]} +\sum_{m=1}^{\big({{N}\atop {p+1}}\big)}{\cal C}_{p+1[lm]}f'_{p+1[m]} =0\   ,\label{recfprime20}
\eeqa
and 
\beqa
\big(\kappa\lambda_n-\Lambda_1\big)f'_{n[l]}+ \sum_{m=1}^{\big({{N}\atop {n-1}}\big)}{\cal B}_{n-1[lm]}f'_{n-1[m]} + \sum_{m=1}^{\big({{N}\atop {n}}\big)}{\cal A}_{n[lm]}f'_{n[m]}=0\qquad \mbox{for}\qquad l=1,...,\Big({{N}\atop {n}}\Big)\   .\label{recfprime2}
\eeqa
\vspace{0.7cm}

\centerline{\bf \large Appendix B: Identical polynomial eigenfunctions for the descendents}
\vspace{3mm}
 
The purpose of this Appendix is to give further support on the fact that higher operators of the form ${\textsf W}_{-k}$, for instance ${\textsf W}_{-1}$, are diagonalized by the polynomial eigenfunctions (\ref{psi}), in agreement with $\big[{\textsf W}_{0},{\textsf W}_{-k}\big]=0$ for all $k=0,1,...$. One has \cite{TriDiag}
\beqa
{\textsf W}_{-1}= -\frac{1}{\rho}\big[{\textsf W}_0,\big[{\textsf W}_0,{\textsf W}_1\big]_q]_{q^{-1}} +
{{\textsf W}_1}\qquad \mbox{and} \qquad {\textsf W}_{2}=-\frac{1}{\rho^*}\big[{\textsf W}_1,\big[{{\textsf W}_1},{{\textsf W}_0}\big]_q]_{q^{-1}}
 + {{\textsf W}_0}\ .\label{defop2}
\eeqa 
Using previous results, suppose $\{{\textsf W}_{0},{\textsf W}_{1}\}$ take the form (\ref{real}) and assume $\phi,{\overline \phi}$ and $d$ solve (\ref{cond}).  Replacing (\ref{real}) in (\ref{defop2}), after simplifications one finds the following action on the infinite dimensional module
\beqa
{\textsf W}_{-1}: && \;\phi_{-1}(z)\eta \;+ \;{\overline \phi}_{-1}(z)\eta^{-1}\;+\;\mu_{-1}(z,z^{-1})I\ ,
\nonumber
\\
{\textsf W}_2: && \;\nu_2(z,z^{-1})\ \label{real2}
\eeqa
with
\beqa
\nu_2(z,z^{-1})&=&\left(1-\frac{(z+z^{-1})^2}{(q^{1/2}+q^{-1/2})^2}\right)\mu(z,z^{-1})\ ,\nonumber\\
\phi_{-1}(z)&=& -\frac{1}{\rho}\Big( \big((q-q^2)z+(q^{-1}-q^{-2})z^{-1}\big)\mu(qz,q^{-1}z^{-1})+\big((q^{-1}-1)z + (1-q)z^{-1}\big)\mu(z,z^{-1})\Big)\phi(z)\ ,\nonumber\\
{\overline \phi}_{-1}(z)&=&-\frac{1}{\rho}\Big( \big((q^{-1}-q^{-2})z+(q-q^2)z^{-1}\big)\mu(q^{-1}z,qz^{-1})+\big((1-q)z + (1-q^{-1})z^{-1}\big)\mu(z,z^{-1})\Big){\overline\phi}(z)\ ,\nonumber\\
\mu_{-1}(z,z^{-1})&=& -\frac{1}{\rho} \Big( \big((q^2-1)z+(q^{-2}-1)z^{-1}\big)\phi(z){\overline\phi}(qz)  +\big((q^{-2}-1)z+(q^{2}-1)z^{-1}\big){\overline\phi}(z){\phi}(q^{-1}z) \ \nonumber\\      
&& \ \ \ + (z+z^{-1})\big((q+q^{-1}-2)(\mu(z,z^{-1})\big)^2\Big)+ z+z^{-1}\ .       \nonumber
\eeqa
The tridiagonal structure (\ref{real2}) is not surprising: similarly to the finite dimensional case\,\footnote{Important discussions with K. Koizumi about this point are aknowledged.}, the action of higher order operators can be written solely in terms of the $q-$difference operators $\eta^{\pm 1}$ and the identity $I$.
The $q-$Sturm-Liouville problem associated with ${\textsf W}_{-1}$ takes the form
\beqa
\phi_{-1}(z)\psi(qz) + {\overline \phi}_{-1}(z)\psi(q^{-1}z) +\mu_{-1}(z,z^{-1})\psi(z) = {\lambda'}\psi(z)\ \label{qdiff3}
\eeqa
with the definitions above. Polynomial solutions can be constructed similarly to the case of ${\textsf W}_0$.
It is however important to notice that, for any $z$, 
\beqa
\frac{{\overline \phi}_{-1}(z)}{\phi_{-1}(z)}=\frac{{\overline \phi}(z)}{\phi(z)}\ ,\label{rap}
\eeqa 
an immediate consequence of (\ref{beta}) and (\ref{cond}). Then, (\ref{rap}) implies that the system of Bethe equations associated with ${\textsf W}_{0}$ is identical to the one for ${\textsf W}_{-1}$. Consequently, up to some overall factors polynomial eigenfunctions of both operators coincide.

\end{document}